\documentclass[manuscript=article]{achemso}
\usepackage{xcolor} 
\usepackage{graphicx} 

\title{Implementation of McMurchie-Davidson algorithm for Gaussian AO integrals suited for SIMD processors.} 

\author{Andrey Asadchev}
\author{Edward F. Valeev}
\email{efv@vt.edu}
\affiliation{Department of Chemistry, Virginia Tech, Blacksburg, VA 24061}

\date{\today}

\usepackage{amsmath,amssymb}
\usepackage{booktabs}
\usepackage[version=4]{mhchem} 
\usepackage{tablefootnote}

\graphicspath{{figures/}}
\usepackage{physics}
\usepackage{multirow}
\usepackage{xcolor}
\usepackage{caption}
\usepackage{subcaption}
\usepackage{xcolor}
\usepackage{bm}
\usepackage{eufrak} 
\usepackage{yfonts} 
\usepackage[title,titletoc]{appendix}
\usepackage{diagbox}
\usepackage{multicol}
\usepackage{mhchem}

\usepackage[numbers,sort&compress,super]{natbib}
\usepackage{natmove}

\usepackage{hyperref} 
\hypersetup{colorlinks=true, citecolor=blue, urlcolor=blue, linkcolor=blue}
\usepackage{cleveref}
  	\crefname{figure}{Figure}{Figures}
  	\crefname{table}{Table}{Tables}
  	\crefname{equation}{Eq.}{Eqs.}
   	\crefname{section}{Section}{Sections}
  	\crefname{subsection}{Section}{Sections}
  	\crefname{subsubsection}{Section}{Sections}
  	\crefname{algorithm}{Algorithm}{Algorithms}

\newcommand{\code}[1]{\texttt{#1}}
\newcommand\vartextvisiblespace[1][.5em]{%
  \makebox[#1]{%
    \kern.07em
    \vrule height.3ex
    \hrulefill
    \vrule height.3ex
    \kern.07em
  }
}
\usepackage{todonotes}

\usepackage[title,titletoc]{appendix}

\SectionNumbersOn

\begin{document}

\begin{abstract}
We report an implementation of the McMurchie-Davidson evaluation scheme for 1- and 2-particle Gaussian AO integrals designed for processors with Single Instruction Multiple Data (SIMD) instruction sets. Like in our recent MD implementation for graphical processing units (GPUs) [{\em J. Chem. Phys.} {\bf 160}, 244109 (2024)], variable-sized batches of shellsets of integrals are evaluated at a time. By optimizing for the floating point instruction throughput rather than minimizing the number of operations, this approach achieves up to 50\% of the theoretical hardware peak FP64 performance for many common SIMD-equipped platforms (AVX2, AVX512, NEON), which translates to speedups of up to 30 over the state-of-the-art one-shellset-at-a-time implementation of Obara-Saika-type schemes in {\tt Libint} for a variety of primitive and contracted integrals. As with our previous work, we rely on the standard C++ programming language --- such as the {\tt std::simd} standard library feature to be included in the 2026 ISO C++ standard --- without any explicit code generation to keep the code base small and portable. The implementation is part of the open source {\tt LibintX} library freely available at \url{https://github.com/ValeevGroup/libintx}.
\end{abstract}

\maketitle

\section{Introduction}

While graphical processing units (GPUs) and other specialized hardware have become the focus of high-performance computing and the key source of computational power on many modern and especially high-end compute platforms, performance-portable CPU kernels will remain a relevant target for the foreseeable future due to several factors.
\begin{enumerate}
\item Although on high-end machines the CPUs provide as little as a few percent of the total floating-point operations (FLOPs), on many platforms (such as a typical laptop, campus-level computing centers, or even some supercomputers) CPUs routinely account for between 50 and 100 \% of simulation-quality (FP64) FLOPs.
\item Some problems may be more suitable for execution on the CPU than the GPU due to the computation task traits, such as where the data resides or the excessive overhead of the data transfer or kernel launch.
\item Recent evolution of the GPU hardware points towards increasing specialization of the GPU cores towards compute patterns dominated by extremely low precision floating point computation and less representative of modern scientific simulation.
\item Despite significant differences, key architectural and programming traits, such as vector execution units and the importance of divergence-free data-parallel computation pattern, are common to both CPUs and GPUs. This presents an opportunity to share the algorithmic and implementation details between CPU and GPU engines, thereby reducing the implementation and maintenance costs.
\item Near-optimal CPU kernels are a proper reference benchmark for the GPU
counterparts that can be used for faithful performance evaluation, constructing a realistic performance model, and identifying further opportunities for optimization. The relatively low performance of the reference code is one of the reasons for spectacular speedups that exceed the relative (GPU-vs-CPU) architectural characteristics; this was the case for our own comparisons of the data-parallel GPU code in \code{LibintX} vs mostly scalar CPU code in \code{Libint},\cite{VRG:asadchev:2023:JPCA,VRG:asadchev:2023:JCTC,VRG:asadchev:2024:JCP} and continue to plague other literature on the Gaussian AO integral evaluation.\cite{VRG:yokogawa:2024:2TISCNWC}
\end{enumerate}

Gaussian AOs\cite{VRG:boys:1950:PRSMPES} are the dominant numerical technology for electronic structure simulation of molecules and have made significant strides towards the condensed phase simulation. Evaluation of operator representation in Gaussian AOs (``Gaussian AO integrals'') accounts for the dominant costs of many types of electronic structure simulation in the Gaussian AO basis, and thus the efficiency of Gaussian integral evaluation is crucial to the efficiency of most common types of molecular electronic structure simulation.
Unfortunately, the majority of Gaussian integral engines for CPUs predate recent advancements in CPU designs such as the emergence of non-x86 HPC-capable architectures (such as ARM), the emergence of wide vector single-instruction multiple-data (SIMD) instruction sets, and AI-driven architectural bias and outright hardware specialization (e.g., matrix execution units on Intel and Apple processors) for matrix operations. For example, our own popular open source engine \code{Libint} achieves
less than 10\% of the peak performance on x86 CPUs with the AVX2 instruction sets and the other popular open-source engine \code{LibCint}
seems to have a similar performance when we compare the reported performance results\cite{VRG:sun:2015:JCC} to {\tt Libint}.

The key challenge of exploiting the wide SIMD vector units of modern CPUs is the lack of a regular source of data parallelism in the Gaussian integral kernels.
Integral engines such as \code{Libcint} that use the Rys quadrature scheme can exploit data parallelism over the quadrature roots while still evaluating one shellset at a time.\cite{VRG:sun:2015:JCC}
A recent report suggested that such strategy is effective only for short vectors (SSE, not AVX) due to the inability to fill the entire SIMD vector with work except for integral classes with high angular momenta.\cite{VRG:sun:2024:JCP}
For auxiliary-function evaluation schemes such as Obara-Saika and McMurchie-Davidson, vectorization over individual integrals in a shell set is not possible due to the irregular arithmetic structure of such compute schemes. Furthermore, the contraction degrees of Gaussian AOs in most practical basis sets are usually too low to be able to populate a whole SIMD vector with primitive shellsets of a single contracted shellset. Therefore, one must consider evaluation of multiple shellsets at a time. Pritchard and Chow explored vectorization over the primitives of the vertical recurrence part of the Obara-Saika-Head-Gordon-Pople (OSHGP) scheme\cite{VRG:obara:1986:JCP,VRG:head-gordon:1988:JCP} when computing multiple shellsets at a time.\cite{VRG:pritchard:2016:JCC}
On an AVX core their implementation achieved speedups of $\sim 2$ over the scalar code and over \code{Libint} for the augmented correlation-consistent triple-zeta basis (with medium contraction degree) and higher speedups for deeply contracted atomic natural orbital triple-zeta basis; unfortunately no results were reported for basis sets with low contraction degrees, such as the Turbomole def2 basis set family.\cite{VRG:weigend:2005:PCCP}

Unfortunately, it is clear that efficient evaluation of integrals on modern CPU cores must abandon the 1 shellset at a time strategy and must follow the multiple shellset at a time strategy that is the only viable route on GPUs.
However, the need for an integral kernel redesign for the CPU presents an opportunity to future proof the CPU code and align the CPU and GPU implementations under the same algorithmic and implementation framework.
In this paper, we report the new implementation of the McMurchie-Davidson scheme\cite{VRG:mcmurchie:1978:JCP} for evaluation of Gaussian AO integrals on modern CPU architectures with SIMD instruction sets in the open source library \code{LibintX}.
To make it easier to adopt the CPU engine and use it alongside the GPU engine, the API of the former is largely identical to the API of the 2-particle integral GPU engine reported previously.\cite{VRG:asadchev:2024:JCP} Another goal was to leverage the algorithmic designs and reuse the code as much as possible between the CPU and GPU engines. Lastly, the key goal was performance portability: the design must be deployable and performant on not only the x86 and ARM architectures dominant in HPC today but also positioned for deployment to future architectures.
Just like the GPU code in \code{LibintX}, only the standard C++ language/library is used without resorting to custom code generation to make the code as transparent and as simple to maintain and evolve as possible. Performance-portable and future-proof vectorization is provided by the
\code{std::experimental::simd} library, a component of the \code{libstdc++} standard C++ library that has recently been accepted into the draft 2026 ISO C++ standard.

The manuscript is structured as follows. \cref{sec:formalism} describes the salient details of the formalism and implementation of the McMurchie-Davidson scheme. \cref{sec:performance} reports the performance assessment of the vectorized implementation relative to the scalar baseline implementation and the scalar reference implementation in \code{Libint}. \cref{sec:summary} summarized our findings.

\section{Formalism}
\label{sec:formalism}

\subsection{McMurchie-Davidson Scheme}

We use the same notation established in our earlier work;\cite{VRG:asadchev:2023:JPCA,VRG:asadchev:2024:JCP} it is largely based on that of Obara and Saika.\cite{VRG:obara:1986:JCP}

\begin{align}
    \phi_\mathbf{a} (\mathbf{r}) \equiv x_A^{a_x} y_A^{a_y} z_A^{a_z} \exp(-\zeta_a r_A^2),
\end{align}
denotes an uncontracted primitive Cartesian Gaussian with exponent $\zeta_a \in\mathbb{R}^+$ and non-negative Cartesian ``quanta'' $\mathbf{a} \equiv \{a_x, a_y, a_z \}, a_i \in \mathbb{Z}$ centered at $\mathbf{A} \equiv \{A_x, A_y, A_z\}, A_i\in \mathbb{R}$,
where $\mathbf{r}_A \equiv \{x_A, y_A, z_A\}, x_A \equiv x - A_x$, etc.
The sum of Cartesian quanta, $l_\mathbf{a} \equiv a_x + a_y + a_z \geq 0$, is colloquially referred to as the ``angular momentum'' of a Gaussian.

The objective of this work is the evaluation of all matrix elements of nonrelativistic Hamiltonian and the basis metric. This involves evaluation
of several types of 1- and 2-particle integrals.
The 1-particle integrals,
\begin{align}
    [\mathbf{a}\mathbf{b}]_{\hat{O}} \equiv & \int \phi_a(\mathbf{r}_1)
    \hat{O} \phi_b(\mathbf{r}_1) \, \mathrm{d}\mathbf{r}_1,
\end{align}
include $\hat{O} = 1$ (overlap), $\hat{O} \equiv \hat{T} = - \nabla^2/2$ (kinetic energy), and $\hat{O} \equiv \hat{\phi} = - \sum_I q_I / r_I$ (Coulomb potential due to point charges $q_I$ at $\mathbf{I} = \{I_x, I_y, I_z\}$, also known as nuclear attraction).
2-particle integrals include the 4-center Coulomb integrals
\begin{align}
\label{eq:def-abcd}
[\mathbf{a} \mathbf{b}|\mathbf{c} \mathbf{d}] \equiv & \iint \frac{\phi_a(\mathbf{r}_1) \phi_b(\mathbf{r}_1) \phi_c(\mathbf{r}_2) \phi_d(\mathbf{r}_2)}{\vert\mathbf{r}_1 - \mathbf{r}_2\vert} \, \mathrm{d}\mathbf{r}_1 \, \mathrm{d}\mathbf{r}_2 ,
\end{align}
as well as the 3-center variety
\begin{align}
\label{eq:def-acd}
[\mathbf{a}|\mathbf{c} \mathbf{d}] \equiv & \iint \frac{\phi_a(\mathbf{r}_1) \phi_c(\mathbf{r}_2) \phi_d(\mathbf{r}_2)}{\vert\mathbf{r}_1 - \mathbf{r}_2\vert} \, \mathrm{d}\mathbf{r}_1 \, \mathrm{d}\mathbf{r}_2
\end{align}
needed for various approximations to the 2-particle Coulomb interaction (density fitting\cite{VRG:whitten:1973:JCP,VRG:baerends:1973:CP,VRG:dunlap:2000:PCCP} and others)\cite{VRG:hohenstein:2012:JCPa,VRG:pierce:2021:JCTC} and for computing empirical atomic potentials in the context of guess orbital construction.\cite{VRG:lehtola:2020:JCP}
The 1- and 2-particle integrals will also be generically denoted as $[\mathrm{bra}]_{\hat{O}}$ and $[\mathrm{bra}\vert\mathrm{ket}]$, respectively.

The objective of this work is efficient evaluation
of integrals over {\em contracted} Gaussian AOs,
\begin{align}
    \phi_\mathbf{a} (\mathbf{r}) \equiv x_A^{a_x} y_A^{a_y} z_A^{a_z} \sum_{k}^{K_a} c_k \exp(-\zeta_{ak} r_A^2),
\end{align}
with $K_a \geq 1$ denoted by contraction {\em degree}.
In the standard McMurchie-Davidson (MD) scheme\cite{VRG:mcmurchie:1978:JCP} used here the integrals are computed in the primitive basis and accumulated into the contracted integrals; this is different from the Head-Gordon-Pople refinement\cite{VRG:head-gordon:1988:JCP} of the Obara-Saika scheme\cite{VRG:obara:1986:JCP} implemented in \code{Libint} where nontrivial work is performed in the contracted basis.
The contracted integrals will be denoted by the use of parentheses instead of square brackets. The total contraction degree of an integral is the product of the contraction degrees of the constituent AOs.
For 2-particle integral $[\text{bra}\vert\text{ket}]$ the contraction degrees of the bra and ket contraction degrees will be denoted as $K_\text{bra}$ and $K_\text{ket}$, respectively; clearly, $K=K_\text{bra} K_\text{ket}$.

The MD approach uses Hermite Gaussians,
\begin{align}
    \Lambda_\mathbf{\tilde{a}} ({\bf r}) \equiv \left(\frac{\partial}{\partial x_A}\right)^{\tilde{a}_x} \left(\frac{\partial}{\partial y_A}\right)^{\tilde{a}_y} \left(\frac{\partial}{\partial z_A}\right)^{\tilde{a}_z} \exp(-\zeta_a r_A^2),
\end{align}
to (exactly) expand primitive Cartesian Gaussians and their binary products:
\begin{align}
\label{eq:cart2herm-1}
    \phi_\mathbf{a}({\bf r}) = & \sum_{\tilde{a}_x=0}^{\tilde{p}_x \leq a_x} E_{a_x}^{\tilde{p}_x}  \sum_{\tilde{p}_y=0}^{\tilde{p}_y \leq a_y} E_{a_y}^{\tilde{p}_y} \sum_{\tilde{p}_z=0}^{\tilde{p}_z \leq a_z} E_{a_z}^{\tilde{p}_z} \Lambda_\mathbf{\tilde{p}}(\mathbf{r}) \equiv \sum_\mathbf{\tilde{p}} E_\mathbf{a}^\mathbf{\tilde{p}} \Lambda_\mathbf{\tilde{p}}(\mathbf{r}), \\
\label{eq:cart2herm-2}
    \phi_\mathbf{a}({\bf r})\phi_\mathbf{b}({\bf r}) = & \sum_{\tilde{p}_x=0}^{\tilde{p}_x \leq a_x+b_x} \left(E_x\right)_{a_x b_x}^{\tilde{p}_x}  \sum_{\tilde{p}_y=0}^{\tilde{p}_y \leq a_y + b_y} \left(E_y\right)_{a_y b_y}^{\tilde{p}_y} \sum_{\tilde{p}_z=0}^{\tilde{p}_z \leq a_z + b_z} \left(E_z\right)_{a_z b_z}^{\tilde{p}_z} \Lambda_\mathbf{\tilde{p}}(\mathbf{r}) \equiv \sum_\mathbf{\tilde{p}} E_{\mathbf{a} \mathbf{b}}^\mathbf{\tilde{p}} \Lambda_\mathbf{\tilde{p}}(\mathbf{r}) ,
\end{align}
with Hermite Gaussian exponent and origin for the 1- and 2-center bra given by
$\zeta_p \equiv \zeta_a$, $\mathbf{P} \equiv \mathbf{A}$ and
$\zeta_p \equiv \zeta_a + \zeta_b$, $\mathbf{P} \equiv \frac{\zeta_a \mathbf{A} + \zeta_b \mathbf{B}}{\zeta_a + \zeta_b}$, respectively.
Hermite Gaussian used to expand the 2-center ket will be denoted by $\tilde{\mathbf{q}}$, with parameters $\zeta_q \equiv \zeta_c + \zeta_d$, $\mathbf{Q} \equiv \frac{\zeta_c \mathbf{C} + \zeta_d \mathbf{D}}{\zeta_c + \zeta_d}$. The tilde over Cartesian quanta, such as $\tilde{\mathbf{a}} \equiv \{\tilde{a}_x, \tilde{a}_y, \tilde{a}_y\}, \tilde{a}_i \in \mathbb{Z}$, will be used instead of the tilde-free counterparts $\tilde{\mathbf{a}}$ to distinguish Hermite Gaussians from their Cartesian counterparts; this will be especially useful in quantities involving both types of Gaussians.

The coefficients of Hermite Gaussians in \cref{eq:cart2herm-1,eq:cart2herm-2} factorize along the Cartesian axes: $E_\mathbf{a}^\mathbf{\tilde{p}} \equiv \prod_{i={x,y,z}} E_{a_i}^{\tilde{p}_i} $, and $E_\mathbf{a b}^\mathbf{\tilde{p}} \equiv \prod_{i={x,y,z}}\left(E_i\right)_{a_i b_i}^{\tilde{p}_i} $. They are evaluated straightforwardly by recursion\cite{VRG:mcmurchie:1978:JCP}:
\begin{align}
\label{eq:E1rr}
    \left(E_x\right)_{a_x+1}^{\tilde{p}_x} = & \frac{1}{2\zeta_a} \left(E_x\right)_{a_x}^{\tilde{p}_x - 1} + (\tilde{p}_x + 1) \left(E_x\right)_{a_x}^{\tilde{p}_x + 1} \\
\label{eq:E2rr1}
    \left(E_x\right)_{a_x+1 \, b_x}^{\tilde{p}_x} = & \frac{1}{2\zeta_p} \left(E_x\right)_{a_x \, b_x}^{\tilde{p}_x - 1} + (P_x - A_x) \left(E_x\right)_{a_x b_x}^{\tilde{p}_x} + (\tilde{p}_x + 1) \left(E_x\right)_{a_x b_x}^{\tilde{p}_x + 1} \\
\label{eq:E2rr2}
        \left(E_x\right)_{a_x \, b_x + 1}^{\tilde{p}_x} = & \frac{1}{2\zeta_p} \left(E_x\right)_{a_x \, b_x}^{\tilde{p}_x - 1} + (P_x - B_x) \left(E_x\right)_{a_x b_x}^{\tilde{p}_x} + (\tilde{p}_x + 1) \left(E_x\right)_{a_x b_x}^{\tilde{p}_x + 1};
\end{align}

from these relations the other bra and the ket counterparts can be obtained straightforwardly. Recurrence \cref{eq:E1rr} is bootstrapped by definitions

\begin{align}
\label{eq:E1-00}
\left(E_i\right)_{0 }^{0} = 1 & \\
\label{eq:E1-ap}
E_{a_i}^{\tilde{p}_i} = 0 & , \quad \tilde{p}_i < 0 \textrm{ or } \tilde{p}_i > a_i \text{  or  }
(\tilde{p}_i + a_i) \bmod{2} \ne 0
\end{align}

Similarly, recurrences \cref{eq:E2rr1,eq:E2rr2} are bootstrapped by
\begin{align}
\label{eq:E2-000}
\left(E_i\right)_{0 0 }^{0} = & \exp\left(-\zeta_a \zeta_b \left(A_i-B_i\right)^2/\left(\zeta_a + \zeta_b\right)\right),\\
\label{eq:E2-abp} E_{a_i b_i}^{\tilde{p}_i} = & \, 0, \quad
  \tilde{p}_i < 0 \textrm{ or }
  \tilde{p}_i > a_i + b_i \text{  or  }
  (\tilde{p}_i + a_i + b_i) \bmod{2} \ne 0
\end{align}

The {\em terminal} $E$ coefficients corresponding to the maximum values of $\tilde{p}_i$ ($\tilde{p}_i = a_i$ and $\tilde{p}_i = a_i + b_i$ for the 1- and 2-center cases, respectively), take simple form:
\begin{align}
  \label{eq:E1-aa}
E_{a_i}^{a_i} = & \frac{1}{\left(2\zeta_a\right)^{a_i}} (E_i)^0_{0},\\
  \label{eq:E2-abab}
E_{a_i b_i}^{a_i+b_i} = & \frac{1}{\left(2\zeta_p\right)^{a_i+b_i}} (E_i)^0_{00}.
\end{align}
These permit several algorithmic improvements, detailed in \cref{sec:algorithmic-improvements}.

Since most of the time we are interested in evaluation of integrals over (real) {\em solid harmonic} Gaussian AOs, and for $l\geq 2$ solid harmonics are less numerous than Cartesians, matrices $E$ are first contracted with the Cartesian-to-solids coefficient matrices\cite{VRG:schlegel:1995:IJQC} to produce matrices $H$ that transform from (primitive) Hermite Gaussians to (primitive) real solid Gaussians directly:
\begin{align}
\label{eq:solid2herm-1}
    H_\mathbf{a}^{\mathbf{\tilde{p}}} = & \sum_{\mathbf{\tilde{p}}} C_{l_a m_a}^{\mathbf{a}} E^{\mathbf{\tilde{p}}}_{\mathbf{a}}, \\
\label{eq:solid2herm-2}
    H_\mathbf{a b}^{\mathbf{\tilde{p}}} = & \sum_{\mathbf{a}} C_{l_a m_a}^{\mathbf{a}} \sum_{\mathbf{b}} C_{l_b m_b}^{\mathbf{b}} E^{\mathbf{\tilde{p}}}_{\mathbf{a b}},
\end{align}
where $C_{l_a m_a}^{\mathbf{a}}$ is the coefficient of Cartesian Gaussian $\phi_\mathbf{a}$ in solid harmonic Gaussian $\phi_a$ with angular momentum quanta $l_a$ and $m_a$; for the detailed definitions see  Ref. \citenum{VRG:schlegel:1995:IJQC}.

{\em Shells} are groups of solid harmonic and Hermite Gaussians that share exponents, origin, contraction coefficients, and orbital quantum numbers. There are $2 l+1$ and $(l+1)(l+2)(l+3)/6$ solid harmonic and Hermite Gaussians in a shell of angular momentum $l$, respectively. Shells of angular momenta \{0, 1, 2, 3, 4, 5, 6\} will be denoted by \{s, p, d, f, g, h, i\}, or by their angular momentum itself. E.g., [21$\vert$43] and [dp$\vert$gf] will denote a {\em class} of integrals representing any {\em shellset} of 4-center integrals with d, p, g, and f shells on each respective center.

Evaluation of the overlap and kinetic energy integrals follows the standard MD scheme in which they are reduced to 1-dimensional integrals.\cite{VRG:mcmurchie:1978:JCP,VRG:helgaker:2000:} Thus here we only focus on the Coulomb 1- and 2-particle integrals.
The use of \cref{eq:cart2herm-1,eq:cart2herm-2,eq:solid2herm-1,eq:solid2herm-2} allows to express Coulomb 1- and 2-particle AO integrals over contracted solid harmonic Gaussians as linear combinations
of the integrals involving primitive Hermite Gaussians only:
\begin{align}
\label{eq:p}
[\mathbf{\tilde{p}}]_{\hat{\phi}} \equiv & \,
\int
\hat{\phi} \Lambda_\mathbf{\tilde{p}}(\mathbf{r}_1) \, \mathrm{d}\mathbf{r}_1 \\
[\mathbf{\tilde{p}} | \mathbf{\tilde{q}}] \equiv & \,
\label{eq:pq}
\iint
\frac{\Lambda_\mathbf{\tilde{p}}(\mathbf{r}_1) \Lambda_\mathbf{\tilde{q}}(\mathbf{r}_2)}{\vert\mathbf{r}_1 - \mathbf{r}_2\vert} \, \mathrm{d}\mathbf{r}_1 \,\mathrm{d}\mathbf{r}_2.
\end{align}
Both can be obtained
from the auxiliary integrals,\footnote{Note that $\mathbf{r}$ refers to Cartesian coordinates of particles whereas $\tilde{\mathbf{r}}$ denotes Cartesian quanta of Hermite Gaussians.}
\begin{align}
\label{eq:rI}
[\mathbf{\tilde{r}}]_{\mathbf{I}}^{(m)} \equiv \left(\frac{\partial}{\partial P_x}\right)^{\tilde{r}_x} \left(\frac{\partial}{\partial P_y}\right)^{\tilde{r}_y} \left(\frac{\partial}{\partial P_z}\right)^{\tilde{r}_z} [\mathbf{0}]_{\mathbf{I}}^{(m)},\\
\label{eq:r}
[\mathbf{\tilde{r}}]^{(m)} \equiv \left(\frac{\partial}{\partial P_x}\right)^{\tilde{r}_x} \left(\frac{\partial}{\partial P_y}\right)^{\tilde{r}_y} \left(\frac{\partial}{\partial P_z}\right)^{\tilde{r}_z} [\mathbf{0}]^{(m)},
\end{align}
via
\begin{align}
\label{eq:p-from-rI}
[\mathbf{\tilde{p}}]_{\hat{\phi}} \equiv  - \sum_I q_I [\mathbf{\tilde{p}}]_{\mathbf{I}}^{(0)}, \\
\label{eq:pq-from-r}
[\mathbf{\tilde{p}} | \mathbf{\tilde{q}}] \equiv (-1)^{l_\mathbf{\tilde{q}}} [\mathbf{\tilde{p}}+\mathbf{\tilde{q}}]^{(0)}.
\end{align}
$[\mathbf{0}]_{\mathbf{I}}^{(m)}$ and $[\mathbf{0}]^{(m)}$ are related to the Boys function $F_m(x)$ (or similar quantities for non-Coulomb integrals\cite{VRG:ahlrichs:2006:PCCP}):
\begin{align}
\label{eq:0Im}
[\mathbf{0}]_{\mathbf{I}}^{(m)} \equiv \, & \frac{2 \pi}{\zeta_p} (-2 \zeta_p)^m  F_m(\zeta_p |\mathbf{P}-\mathbf{I}|^2) , \\
\label{eq:0m}
[\mathbf{0}]^{(m)} \equiv \, & \frac{2 \pi^{5/2}}{\zeta_p \zeta_q \sqrt{\zeta_p+\zeta_q}} (-2 \rho)^m F_m(\rho |\mathbf{P}-\mathbf{Q}|^2) , \\
F_m(x) \equiv \, & \int_0^1 \, \mathrm{d}y \, y^{2m} \exp(-x y^2), \label{eq:boys} \\
\rho \equiv \, & \frac{\zeta_p \zeta_q}{ \zeta_p + \zeta_q }.
\end{align}
The auxiliary integrals are evaluated recursively,
\begin{align}
\label{eq:rI-rr}
[\mathbf{\tilde{r}}+\mathbf{1}_i]_{\mathbf{I}}^{(m)} = & \tilde{r}_i [\mathbf{\tilde{r}}-\mathbf{1}_i]_{\mathbf{I}}^{(m+1)} + \left(P_i - I_i\right) [\mathbf{\tilde{r}}]_{\mathbf{I}}^{(m+1)},\\
\label{eq:r-rr}
[\mathbf{\tilde{r}}+\mathbf{1}_i]^{(m)} = & \tilde{r}_i [\mathbf{\tilde{r}}-\mathbf{1}_i]^{(m+1)} + \left(P_i - Q_i\right) [\mathbf{\tilde{r}}]^{(m+1)},
\end{align}
starting from $[\mathbf{0}]_{\mathbf{I}}^{(m)}$ and $[\mathbf{0}]^{(m)}$, respectively.

Transformation of 2-particle integrals from Hermite to AO basis assumes that the ket has higher $K$ and lower $l$ than the bra, hence the former is transformed first:
\begin{align}
\label{eq:ket-from-hermite}
[\tilde{\mathbf{p}}\vert\mathrm{ket}] = & \sum_{\tilde{\mathbf{q}}} [\tilde{\mathbf{p}}\vert \tilde{\mathbf{q}}] H_\mathrm{ket}^{\tilde{\mathbf{q}}}, \\
\label{eq:bra-from-hermite}
[\mathrm{bra}\vert\mathrm{ket}] = & \sum_{\tilde{\mathbf{p}}} [\tilde{\mathbf{p}}\vert \mathrm{ket}] H_\mathrm{bra}^{\tilde{\mathbf{p}}}.
\end{align}
For 1-particle Coulomb integrals there is only the bra to transform:
\begin{align}
\label{eq:bra-from-hermite-1body}
[\mathrm{bra}]_{\hat{\phi}} = & \sum_{\tilde{\mathbf{p}}} [\tilde{\mathbf{p}}]_{\hat{\phi}} H_\mathrm{bra}^{\tilde{\mathbf{p}}}.
\end{align}

The complete sequences of transformations involved in the MD scheme for the evaluation of {\em primitive} Coulomb 1-particle and 2-particle integrals are
\begin{align}
\label{eq:md-1body}
[\mathbf{0}]_{\mathbf{I}}^{(m)} &
\overset{\mathrm{\cref{eq:rI-rr}}}{\to} [\tilde{\mathbf{r}}]_{\mathbf{I}}^{(0)}
\overset{\mathrm{\cref{eq:p-from-rI}}}{\to} [\tilde{\mathbf{p}}]_{\hat{\phi}} \overset{\mathrm{\cref{eq:bra-from-hermite-1body}}}{\to} [\mathrm{bra}]_{\hat{\phi}} \\
\label{eq:md-2body}
[\mathbf{0}]^{(m)} &
\overset{\mathrm{\cref{eq:r-rr}}}{\to} [\tilde{\mathbf{r}}]^{(0)}
\overset{\mathrm{\cref{eq:pq-from-r}}}{\to} [\tilde{\mathbf{p}}\vert\tilde{\mathbf{q}}]
\overset{\mathrm{\cref{eq:ket-from-hermite}}}{\to} [\tilde{\mathbf{p}}\vert\mathrm{ket}] \overset{\mathrm{\cref{eq:bra-from-hermite}}}{\to} [\mathrm{bra}\vert\mathrm{ket}]
\end{align}
The required intermediate integrals satify the following constraints: $m\in[0,L]$, $l_{\tilde{\mathbf{r}}} \in [0,L]$, $l_{\tilde{\mathbf{p}} }\in [0,l_\text{bra}]$, and $ l_{\tilde{\mathbf{q}}} \in [0,l_\text{ket}]$, where $l_\text{bra}$, $l_\text{ket}$, and $L = l_\text{bra} + l_\text{ket}$, are the total angular momenta of the bra, ket, and whole integral, respectively

A simple roofline-style performance model for the latter was constructed in Ref. \citenum{VRG:asadchev:2024:JCP}; the key takeaway is that for the percentage of FLOPs  spent on the transformation from Hermite to AO basis grows rapidly with the angular momentum, thus, as pointed out by Neese in his pioneering work on the SHARK engine,\cite{VRG:neese:2022:JCC} optimizing the transformation by expressing it as a GEMM kernel makes sense. As we showed\cite{VRG:asadchev:2024:JCP} lower angular momenta can also be efficiently evaluated in matrix form using custom inline small matrix multiplication (matmul) kernel rather than BLAS or CUTLASS GEMM.

\subsection{Algorithmic Improvements}
\label{sec:algorithmic-improvements}

Several algorithmic improvements to the original MD scheme are possible. Although straightforward and obvious, we have not seen them mentioned in the literature, so they are emphasised here.

\subsubsection{Coulomb 1-particle integrals}
\label{sec:algorithmic-improvements-2c1e}

Transformation from Hermite to the AO basis does not depend on nuclear coordinates and charges, and thus can be done outside of summation over centers.

\begin{align}
\label{eq:bra-as-sum-over-charges}
[\mathrm{bra}]_{\hat{\phi}} = - \sum_I q_I [\mathrm{bra}]_\mathbf{I} = - \sum_I q_I
\sum_{\tilde{\mathbf{p}}}
[\tilde{\mathbf{p}}]_{\mathbf{I}} H_\mathrm{bra}^{\tilde{\mathbf{p}}}  &= - \sum_{\tilde{\mathbf{p}}} H_\mathrm{bra}^{\tilde{\mathbf{p}}} ( \sum_I q_I [\tilde{\mathbf{p}}]_{\mathbf{I}})
\end{align}
Although obvious in retrospect,
this optimization does not seem to be well known as even very recent papers appear to use the naive approach\cite{VRG:yokogawa:2024:2TISCNWC}.
Although an extension of the Coulomb 1-particle OS VRR with early summation over charges is possible,
albeit not as simple, it has not been reported to the best of our knowledge and does not appear to offer any advantage over MD.

\subsubsection{Early contraction of the $E_{\mathbf{a} \mathbf{b}}^{\mathbf{a}+\mathbf{b}}$ contribution}
\label{sec:algorithmic-improvements-Eab}

The simple form of \cref{eq:E2-abab} allows us to express the contributions to Cartesian AOs from Hermite Gaussians with the terminal (maximal) Hermite quantum numbers in a compact form, e.g., \cref{eq:cart2herm-2} becomes:
\begin{align}
\label{eq:cart2herm-2-opt}
\phi_\mathbf{a}({\bf r})\phi_\mathbf{b}({\bf r}) = &
\frac{1}{\left(2\zeta_p\right)^{l_\mathbf{a}+l_\mathbf{b}}} \Lambda_\mathbf{\tilde{0}}(\mathbf{r}) +
\sum_\mathbf{\tilde{p} \neq a + b} E_\mathbf{ab}^\mathbf{\tilde{p}} \Lambda_\mathbf{\tilde{p}}(\mathbf{r}).
\end{align}
Eliminating the explicit appearance of Hermite Gaussians with terminal quantum numbers \cref{eq:cart2herm-2-opt} allows us to reduce the range of quantum numbers of the auxiliary and Hermite integrals and their footprint.
Applying this optimization to the Hermite-to-Cartesian ket transformation in the context of \cref{eq:ket-from-hermite}, we obtain the following:
\begin{align}
\label{eq:ket-from-hermite-opt}
[\tilde{\mathbf{p}}\vert\mathrm{ket}] = &
\sum_\mathbf{cd} C^\mathbf{cd}_\mathrm{ket} \frac{1}{\left(2\zeta_q\right)^{l_\mathbf{c}+l_\mathbf{d}}} [\tilde{\mathbf{p}}\vert \tilde{\mathbf{0}}] +
\sum_{\mathbf{\tilde{q} \neq c+d}} [\tilde{\mathbf{p}}\vert \tilde{\mathbf{q}}] H_\mathrm{ket}^{\tilde{\mathbf{q}}} \nonumber \\
\overset{\text{\cref{eq:pq-from-r}}}{=} & \sum_\mathbf{cd} C^\mathbf{cd}_\mathrm{ket} \frac{1}{\left(2\zeta_q\right)^{l_\mathbf{c}+l_\mathbf{d}}} [\tilde{\mathbf{p}}]^{(0)} +
\sum_{\mathbf{\tilde{q} \neq c+d}} [\tilde{\mathbf{p}}\vert \tilde{\mathbf{q}}] H_\mathrm{ket}^{\tilde{\mathbf{q}}},
\end{align}
where $\sum_\mathbf{cd} C^\mathbf{cd}_\mathrm{ket}$ denotes the Cartesian-to-solid-harmonics transformation of the ket.
\cref{eq:ket-from-hermite-opt} was previously exploited in our MD GPU integral engine\cite{VRG:asadchev:2024:JCP} to reduce the memory footprint of the auxiliary and Hermite integrals and the number of operations. This is especially significant for lower angular momenta.  For example, for $l_\text{ket}=2$ (such as integrals with $\vert\text{pp}]$ or $\vert\text{ds}]$ kets) this technique allows to compute directly the contributions to \cref{eq:ket-from-hermite-opt} from 6 Hermite Gaussians with $l_{\tilde{\bf q}}=2$ which account for 60\% of the total of 10 Hermite Gaussians in the d Hermite shell.

It is straightforward to use \cref{eq:ket-from-hermite-opt} to compute the contributions of this term to the target integral with {\em contracted} ket:
\begin{align}
\label{eq:contracted-ket-from-hermite-opt}
  [\tilde{\mathbf{p}}\vert\mathrm{ket}] =  & \sum_\mathbf{cd} C^\mathbf{cd}_\mathrm{ket} \sum_{k_\text{ket}} \frac{1}{\left(2\zeta_q\right)^{l_\mathbf{c}+l_\mathbf{d}}} [\tilde{\mathbf{p}}]^{(0)} +
\mathrm{contributions\ from\ \mathbf{\tilde{q} \neq c+d}},
\end{align}
where summation over $k_\text{ket}$ denotes contraction over ket primitives. By allowing to compute contributions to the integrals with contracted ket directly from the auxiliary (1-index) integrals \cref{eq:contracted-ket-from-hermite-opt} can be viewed as employing partial early contraction.
This approach offers little to no improvement for integrals with s and p functions only.  For other classes of integrals the performance improvement is significant due to fewer operations overall and simpler inner kernel.

\subsubsection{Coulomb 2-particle 3-center integrals}
\label{sec:algorithmic-improvements-3c2e}

Evaluation of \cref{eq:bra-from-hermite} for the case of 1-center bra in general requires only integrals with $\tilde{p}_i$ of the same parity as $a_i$ due to \cref{eq:E1-ap}.

In the case of {\em solid harmonic} bra (rather than Cartesian), it is sufficient to include integrals with $\tilde{p}_i = a_i$, that is, {\em only} the aforementioned terminal term.
This is due to the fact that the solid harmonic of the angular momentum $l$ is a linear combination of monomials $x^{a_x} y^{a_y} z^{a_z}$ with $a_x + a_y + a_z = l$, and such monomials are only present in the Hermite Gaussian $\Lambda_{\tilde{\mathbf{p}}}$ with $\tilde{\mathbf{p}} = \mathbf{a}$.

Note that the required range of $m$ in $[0]^{(m)}$ matches the range required by the Ahlrichs' improvement of the OS scheme for 3-center 2-particle integrals over solid harmonic Gaussians.\cite{VRG:ahlrichs:2004:PCCP} The match is not coincidental as both Ahlrichs' and our optimizations omit the intermediate Cartesian/Hermite integrals that produce contributions to the target $[\mathbf{a}|\mathbf{cd}]$ integral with the overly slow multipolar decay, and hence such contributions are filtered out by the transformation to the solid harmonic basis.

This optimization reduces the {\em entire} hermite-to-AO transformation of the auxiliary bra
to just that of the terminal term.  Moreover if bra is contracted, early bra contraction reduces to
contracting $[\mathbf{0}]^{(m)}$ integrals rather than 1-index $[\tilde{\mathbf{p}}]^{(0)}$  integrals
as in \cref{eq:contracted-ket-from-hermite-opt}.
However, since the density fitting AOs are usually uncontracted, this will result in rather modest actual improvement.
This idea has been described by Knizia et al.\cite{VRG:peels:2020:JCTC} for {\em 2-center} 2-particle integrals over solid harmonic Gaussians but generalization to the 3-center case was not reported.

\section{Implementation}
\label{sec:implementation}

\subsection{Basics}
\label{sec:implementation-basics}

The new SIMD-capable MD implementation for standard 1-particle and up to 4-center 2-particle integrals was implemented in the \code{LibintX} library, freely available at \textbf{\url{github.com:ValeevGroup/LibintX}} (branch \code{feature/md/simd}) under the Lesser GNU Public License (version 3).
As in our previous\cite{VRG:asadchev:2023:JCTC,VRG:asadchev:2023:JPCA,VRG:williams-young:2023:JCP,VRG:asadchev:2024:JCP} work the library is written in standard C++; there is no custom code generator.
Modern metaprogramming facilities of C++ allow to reach high performance while keeping the source code compact (fewer than 1,500 lines in total). Similar to the GPU implementation\cite{VRG:asadchev:2024:JCP} the integral kernel compilation is broken into multiple object targets.
Special attention was paid to ensure that time and memory
requirements for compilation remain reasonable even for high angular momenta.
For example, compilation of the library configured to support \{3,4\}-center integrals with $l=6$ AOs takes \{5,15\} minutes using the LLVM Clang C++ compiler on four i7 cores, using $<$1GB of memory per translation unit.

There are no external dependencies beyond the standard C++ library.
If SIMD is enabled, the recently standardized \code{std::simd} library is needed, available in the GNU libstc++ 11 and above.

Simple compile-time
arithmetic is used to tile small matrix kernels for best register usage.  If compiled with Intel Math Kernel Library (MKL) BLAS then some matrix kernels are instantiated via MKL's Just-In-Time (JIT) interface.

The CPU and GPU integral engines share the same interface via virtual base class.
Both produce integrals in the same $[ij,ab,cd,kl]$ column-major layout (with $ij$ and $kl$ the bra and ket shell-pair indices and $ab$ and $cd$ the bra and ket AO pairs; see Ref. \citenum{VRG:asadchev:2024:JCP} for the layout discussion of the GPU implementation).
Internally, integrals are computed in small batches (multiples of the SIMD vector size) and the CPU side has an interface to ``consume'' (process) those small batches on the fly rather than populating the entire integral array.  Screening is likewise done on a batch level with assumption that bra (and to lesser effect ket) indices are sorted in (any) order that has blocek sparsity.

\subsection{Specifics}
\label{sec:implementation-specifics}

\subsubsection{Boys functions}
\label{sec:implementation-boys}

Fast vectorized evaluation of the Boys function (\cref{eq:boys}) is
critical to the overall performance of Coulomb integrals (including nuclear attraction),
especially for low angular momenta, where it accounts for a large
percentage of the total cost.

The Boys function can be evaluated via upward recursion
\begin{align}
  F_0(x)     = & \sqrt{\frac{\pi}{x}}\erf\left(\sqrt{x}\right) \\
  F_{m+1}(x) = & \frac{(2m+1)F_m(x)}{2x} - \frac{e^{-x}}{2x}
\end{align}
With $x > 36$ equations simplify to the asymptotic form accurate to the 64-bit floating-point epsilon:
\begin{align}
  \erf\left(\sqrt{x}\right)\approx 1 \\
  \frac{e^{-x}}{2x} \approx 0.
\end{align}
For $L < 9$ and $x > 36$ the upward recursion is stable enough, but in the general
case the upward asymptotic recursion is stable when $x > 118$.  For other values of $L$ and $x$ the Boys function is evaluated using the 7th-order Chebyshev interpolation also used in \code{Libint}.  Alternatives with downward recursion
did not offer any performance improvement due to the high cost of {\tt exp} on all tested CPUs, but may be a better option for GPUs.

The Boys function evaluation for scalar (non-SIMD) argument types is also partially vectorized. Namely, the Chebyshev interpolation evaluation branch is vectorized internally (this is also true for the Boys engine implementation in \code{Libint} on x86-64 platforms). For portability this is implemented using 128-bit GCC vector extensions, wider vectors or explicit intrinsics did not offer noticeable performance improvement.

In evaluation of the Boys function for SIMD argument types both asymptotic and interpolation branches need to be evaluated with ``invalid'' SIMD lanes masked with zeroes.
With AVX2 instructions it is possible to implement faster interpolation (by roughly $2\times$)
by rearranging the 7th order Chebyshev interpolation into two parts and using vector permutation intrinsics.
Such optimization is not possible with 512-bit instructions as the vector size is {\em equal} to
the number of interpolation coefficients.
In fact, this AVX2 implementation is faster than the generic {\tt std::simd} AVX512 version;
therefore, the AVX2 implementation is used regardless of the vector width.  This is the only
place where architecture-specific code is used; interested readers are referred to the source code.

\subsubsection{2-particle integrals}

Unlike the 3 algorithmic variants that were used to adapt the MD implementation to the complex memory hierarchy of the GPU,\cite{VRG:asadchev:2024:JCP} the single algorithm is used for 2-particle integrals, with the variance being whether the bespoke or generic (BLAS) GEMM is used.

For 3-center 2-particle integrals, the 1-center bra $[X|q]$ is transformed first, followed by ket transformation to $[X|cd]$.
If ket fits into registers (for example, $|$dp] fits into 16 registers), a small inline matrix kernel is used, the BLAS GEMM is used otherwise.
If BLAS GEMM is used, $X$ centers are batched together for better performance.
The bra early contraction optimization (\cref{sec:algorithmic-improvements-3c2e}) is {\em only} used with BLAS GEMM at the moment as it inhibits the compiler from some inline code optimizations.  The partial ket early contraction is not used as it did not produce much improvement.

The 4-center integral is evaluated as $[r] \to [p|cd] \to [ab|cd]$, with vectorization over bra as in the 3-center case. As with the 3-center code, larger Hermite-to-AO transforms are handled with BLAS GEMM, smaller with inline matrix kernel. The partial early contraction optimization (\cref{sec:algorithmic-improvements-Eab}) is always used.
{\em Scalar} bra transformation can be done with GEMM as well, but with SIMD leading dimension is now interleaved vector elements, requiring explicit transpose steps to use GEMM  (either before or after).  This also prevents us from using small matrix libraries like \code{libxsmm}.
Instead, SIMD bra transformation uses another bespoke inline matrix kernel.
Bra will tend to dominate where the ket is uncontracted and is much smaller, e.g., (ii$|$pp), but those
cases are less common.  Notice that in such cases BLAS GEMM is (expectedly) faster than the bespoke kernel.

\section{Performance Assessment}
\label{sec:performance}

The performance of scalar and SIMD-enabled MD implementations in \code{LibintX} was compared against the reference \code{Libint} implementation of the Obara-Saika-Head-Gordon-Pople\cite{VRG:obara:1986:JCP,VRG:head-gordon:1988:JCP} scheme (augmented with the Ahlrichs improvement\cite{VRG:ahlrichs:2004:PCCP} for the 3-center 2-particle integrals).
Platforms with instruction sets
AVX2 (256-bit SIMD vectors), AVX512 (512-bit SIMD vectors)
and NEON (128-bit SIMD vectors) were used (see \cref{tab:platforms}).
The \code{-ffast-math -Ofast -march=native} compiler optimization flags were used throughout.

\begin{table}[ht!]
    \centering
    \begin{tabular}{lllll}
    \hline\hline
        Instruction Set & CPU & C++ Compiler & C++ Library & BLAS Library \\ \hline
        Scalar & i7-9750H (mobile) & \code{gcc} 12 & \code{libstdc++} 12 & Intel MKL 2020 \\
        AVX2 & i7-9750H (mobile) & \code{clang} 15 & \code{libstdc++} 12 & Intel MKL 2020 \\
        AVX512 & Xeon Platinum 8160  & \code{gcc} 14 & \code{libstdc++} 14 &  Intel MKL 2024 \\
        NEON & M2 Max & \code{clang} 17 & \code{libstdc++} 14 &  Apple Accelerate$^a$ \\
             \hline\hline
    \end{tabular}
    \caption{Platforms used to assess the performance of the SIMD-capable MD implementation in \code{LibintX}.}
    \label{tab:platforms}
    $^a$ Provided by Apple Xcode 16.
\end{table}

We will denote by Scalar/SIMD the implementation variants that partition batches of shellsets into statically sized batches of 1/n shellsets, respectively. The scalar code still contains
SIMD instructions, either in the Boys engine, BLAS GEMM kernels, or generated by the compiler autovectorizer.

Test Gaussians were generated randomly to mimic typical basis set values and to randomise Boys computations.  Libintx timings include time to compute both, integrals {\em and} shell pair quantities.   Libint timings were obtained by computing a single braket in a loop same number of times as Libintx counterpart with shell pair data initialized {\em outside} the loop.  In both cases screening was turned off and result buffers were {\em not} commited to memory to avoid memory bandwidth artifacts.  Tests were repeated 3-4 times with best times reported however the variance was negligible.
CPU details are included in supplemental information.

All tests were executed with a single thread for ease of reproducibility.  CPU governor was set for performance and frequence scaling {\em was} enabled.  Since it is not possible to control the number of threads used by the Apple Accelerate library on ARM-based Apple platforms, the reported 1-thread speed-ups for high-angular-momentum integrals that use BLAS GEMM {\em may} be overly optimistic (such a caveat also applies to the published benchmarks of the SHARK engine\cite{VRG:neese:2022:JCC} on Apple platforms).

For a handful of AVX2 tests we also report double precision performance as percentage of the peak theoretical throughput, the latter obtained by multiplying the clock frequency as reported by VTune by 16 (approximately 64 GFlops peak).

The raw data including the number of integrals evaluated per second (compute rates) are reported in Supporting Information.

\subsection{1-particle integrals}

As already mentioned, the overlap and kinetic energy integrals are straightforward to implement and account for insignificant costs in practical computations.
Each shell-pair of contracted operlap/kinetic integrals is assigned to a single SIMD lane.
The observed performance increases with vector length and does not differ much from the \code{Libint} reference; therefore, we do not discuss the performance further here.

\Cref{table:md2} reports the performance of our MD implementation for Coulomb (nuclear attraction) integrals due to the potential of 100 point charges. Significant speedups relative to the reference OSHGP implementation in \code{Libint} were observed, since even the scalar code
is at least $2\times$ faster, due to the early summation of point charges possible in the MD framework (\cref{sec:algorithmic-improvements-2c1e}). The greatest speedups of the scalar code ($>13\times$) are observed for the (s$|$s) integrals which only involve the evaluation of Boys functions.
Since the interpolation branch of the Boys engine is vectorized even in scalar code (see \cref{sec:implementation-boys}), the additional speedup of SIMD code vs scalar code is due to the benefit of vectorization of the asymptotic branch of the Boys engine.
For nonzero $L$ our implementation exhibits substantial (albeit smaller) speedups over the reference OSHGP implementation even using scalar instructions, despite the lower operation count of the latter. For primitive integrals vectorization using AVX2, AVX512, and NEON instruction sets produces speedups over the scalar code of $\times1.6-3.3$ ($\times2.4$ average), $\times1.5-4.0$ ($\times3$ average), and $\times1.1-1.9$ ($\times1.5$ average), respectively. For the contracted integrals vectorization speedups are even higher: $1.7-3.9$ ($2.9$ on average), $1.9-5.7$ ($4$ on average), and $1.1-2.7$ ($1.8$ on average), respectively. Note that the speedups over 2 for contracted (f$|$s) integrals with NEON instruction set are somewhat misleading since these figures are relative to the scalar x86-64 code, not to the scalar ARM code.
Note that increasing speedup with $L$ is again primarily due to the ability to optimize the early summation (\cref{sec:algorithmic-improvements-2c1e}) in MD, which allows one to shift more cost outside of the charge summation loop.


\begin{table}[ht!]
  \scalebox{1}{
\begin{tabular}{l|llll|llll}
\hline\hline
       & \multicolumn{4}{c|}{$K=1$} & \multicolumn{4}{c}{$K=5$}     \\
\hline
$(l_\mathbf{a}|l_\mathbf{b})$ & I1 & I4  & I8 & I2  & I1 & I4   & I8 & I2  \\
\hline
 $(0|0)$ & 13.86  & 12.53 & 15.79  & 10.18 & 13.93  & 22.97  & 22.12  & 18.69 \\
 $(1|0)$ & 11.82  & 19.74 & 17.75  & 13.74 & 9.25   & 17.96  & 17.24  & 11.10 \\
 $(2|0)$ & 5.74   & 9.12  & 10.41  & 7.18  & 4.49   & 12.22  & 14.16  & 8.55  \\
 $(3|0)$ & 2.91   & 8.36  & 9.57   & 5.49  & 2.16   & 7.95   & 10.32  & 5.83  \\
 $(4|0)$ & 3.52   & 10.63 & 11.34  & 6.52  & 2.91   & 10.42  & 12.67  & 6.01  \\
 $(5|0)$ & 2.99   & 9.33  & 10.65  & 4.67  & 2.69   & 10.56  & 14.67  & 4.68  \\
 $(6|0)$ & 2.77   & 9.03  & 10.94  & 4.78  & 2.55   & 9.90   & 14.50  & 4.85  \\
 $(1|1)$ & 6.40   & 11.78 & 12.42  & 9.53  & 4.84   & 13.27  & 15.70  & 10.09 \\
 $(2|1)$ & 3.48   & 7.54  & 9.30   & 5.98  & 2.52   & 7.54   & 9.98   & 5.98  \\
 $(3|1)$ & 2.81   & 6.96  & 8.32   & 3.98  & 2.12   & 7.38   & 9.79   & 4.13  \\
 $(4|1)$ & 4.46   & 12.44 & 13.68  & 6.21  & 3.32   & 10.66  & 14.22  & 5.15  \\
 $(5|1)$ & 4.34   & 10.66 & 13.75  & 6.14  & 3.29   & 8.75   & 12.91  & 5.01  \\
 $(6|1)$ & 3.69   & 9.59  & 13.15  & 5.52  & 2.89   & 8.40   & 13.36  & 4.89  \\
 $(2|2)$ & 3.17   & 7.23  & 8.45   & 4.47  & 2.21   & 6.90   & 9.03   & 3.84  \\
 $(4|2)$ & 5.62   & 13.77 & 20.05  & 7.69  & 3.31   & 8.76   & 13.78  & 5.01  \\
 $(6|2)$ & 4.63   & 11.07 & 16.66  & 6.53  & 2.80   & 7.11   & 11.47  & 4.63  \\
 $(3|3)$ & 3.19   & 7.34  & 9.52   & 3.36  & 2.03   & 5.25   & 8.32   & 2.30  \\
 $(6|3)$ & 5.49   & 11.83 & 19.03  & 7.29  & 2.86   & 6.43   & 10.89  & 4.27  \\
 $(4|4)$ & 7.40   & 18.22 & 27.96  & 10.78 & 3.11   & 8.31   & 14.05  & 5.18  \\
 $(5|5)$ & 10.49  & 19.89 & 33.00  & 18.68 & 3.87   & 7.54   & 13.19  & 6.04  \\
 $(6|6)$ & 12.98  & 21.62 & 28.67  & 25.09 & 4.20   & 7.00   & 10.45  & 6.81  \\
\hline\hline
\end{tabular}
}

  \caption{Speedup of the MD implementation of Coulomb 2-center 1-particle integrals in \code{LibintX} vs the reference OSHGP implementation in \code{Libint}. Coulomb potential generated by 100 point charges.$^a$}
  \label{table:md2}
  ~\\
  $^a$ SIMD instruction sets: I1 $\equiv$ scalar x86-64, I4 $\equiv$ AVX2, I8 $\equiv$ AVX512, I2 $\equiv$ NEON.
\end{table}

\subsection{3-center 2-particle integrals}

\Cref{table:md3} reports the performance of our MD implementation for 3-center 2-particle integrals.
For primitive AOs, the performance of the scalar variant of the new code is substantially superior to that of OS,
by $\times 2.5$ or more, with few exceptions.
Namely, the OSHGPA reference implementation in \code{Libint}
performs well where ket is large and bra is small, e.g. $(1|66)$, especially if ket is contracted (which should not occur except in atomic natural orbital basis sets where deeply contracted high-$l$ AOs are encountered).
SIMD vectorized implementations are always faster than the reference. This is supported by measured FLOP throughputs (averaged over the three cases of contraction depths): evaluation of (p$|$pp) and (d$|$dd) integrals with \code{LibintX} achieved 30\% and 50\% of peak, respectively, on the AVX2 platform, while the corresponding \code{Libint} throughputs were 5\% and 7\%, respectively. For higher angular momenta even higher throughputs ($\sim90\%$) are achieved due to the shift of the bulk of the FLOPs to matrix multiplication. The higher FLOP throughput allows our MD-based implementation to overcome the lower operation count of the reference OSHGPA evaluation scheme.


The additional benefit of SIMD vectorization is significant, e.g. on average a factor of 2 speedup is observed with the AVX2 instructions. Note that decreasing SIMD speedup relative to the scalar code is due to the increasing percentage of the cost accounted by the matrix multiplication kernels (which inherently use the SIMD instruction set even in the scalar code). Matrix multiplication also accounts for the anomalously high speedups obtained with the NEON instruction set for high-$L$ integrals (especially, (i$|$ii)).  The details of Apple Matrix Extensions upon which Apple Accelerate relies
are not disclosed at this time.  It is not clear if AMX execution units are per core or shared between some cores, or whether such performance is sustained in multithreaded environment. 
On the other hand, Intel MKL kernels are confined to execute on a single CPU hardware thread.

Note the increased performance when bra is contracted; for example, compare the performance for (p$|$ii) integrals with $K_\text{bra}=1$ vs $K_\text{bra}=5$ with $K_\text{ket}=10$. This is due to the early bra contraction optimization (\cref{sec:algorithmic-improvements-3c2e}).

\begin{table}[ht!]
  \scalebox{1.0}{
\begin{tabular}{l|llll|llll|llll}
 \hline\hline
 & \multicolumn{4}{c|}{$\{K_\text{bra},K_\text{ket}\} = \{1, 1\}$}  & \multicolumn{4}{c|}{$\{K_\text{bra},K_\text{ket}\} = \{1, 10\}$}     &  \multicolumn{4}{c}{$\{K_\text{bra},K_\text{ket}\} = \{5, 10\}$} \\
 \hline
 $(l_\mathbf{a}|l_\mathbf{c}l_\mathbf{d})$     & I1 & I4   & I8 & I2  & I1 & I4   & I8 & I2 & I1 & I4  & I8 & I2    \\
 \hline 
 $(0|00)$ & 6.21    & 8.38   & 7.98  & 6.06  &  4.49    & 5.54 & 5.78  & 3.97 & 4.11    & 5.25 &  5.38  &  3.36    \\
 $(1|00)$ & 16.29   & 29.86  & 25.01 & 14.53 &  5.85    & 9.81 & 9.75  & 6.16 & 4.93    & 7.68 &  7.97  &  5.46    \\
 $(2|00)$ & 11.37   & 27.38  & 23.16 & 12.52 &  4.05    & 8.86 & 9.39  & 5.38 & 3.30    & 6.91 &  7.80  &  4.74    \\
 $(4|00)$ & 7.79    & 23.86  & 21.92 & 12.05 &  2.73    & 7.41 & 9.24  & 5.03 & 2.26    & 5.79 &  8.01  &  4.48    \\
 $(6|00)$ & 6.86    & 25.74  & 25.20 & 9.53  &  2.30    & 7.54 & 10.68 & 3.98 & 1.81    & 5.60 &  8.73  &  3.56    \\
 $(0|11)$ & 5.70    & 17.80  & 14.88 & 7.25  &  2.01    & 5.69 & 6.14  & 2.32 & 1.67    & 4.49 &  5.14  &  2.09    \\
 $(1|11)$ & 3.89    & 13.48  & 11.64 & 5.35  &  1.26    & 4.19 & 4.36  & 1.62 & 1.02    & 3.27 &  3.61  &  1.40    \\
 $(2|11)$ & 3.87    & 10.38  & 11.66 & 4.24  &  1.04    & 3.25 & 4.17  & 1.35 & 0.79    & 2.42 &  3.42  &  1.12    \\
 $(4|11)$ & 2.73    & 9.80   & 13.93 & 5.80  &  1.12    & 3.92 & 6.59  & 2.71 & 0.97    & 3.27 &  5.83  &  2.40    \\
 $(6|11)$ & 2.79    & 11.09  & 15.17 & 6.76  &  1.43    & 5.49 & 8.92  & 3.84 & 1.33    & 4.95 &  8.38  &  3.50    \\
 $(1|22)$ & 2.81    & 5.22   & 6.66  & 3.02  &  0.99    & 1.71 & 2.50  & 1.04 & 3.20    & 5.27 &  7.28  &  2.38    \\
 $(2|22)$ & 2.93    & 5.41   & 6.60  & 3.08  &  1.00    & 1.81 & 2.45  & 1.12 & 3.58    & 6.18 &  8.02  &  2.77    \\
 $(1|33)$ & 2.01    & 3.20   & 4.86  & 1.97  &  0.48    & 0.85 & 1.57  & 0.68 & 1.73    & 3.06 &  5.49  &  2.00    \\
 $(3|33)$ & 2.23    & 3.42   & 4.45  & 3.05  &  0.66    & 1.07 & 1.50  & 1.06 & 2.63    & 4.08 &  5.89  &  3.11    \\
 $(1|44)$ & 1.62    & 2.03   & 3.95  & 2.56  &  0.34    & 0.49 & 0.95  & 0.77 & 1.28    & 1.80 &  3.33  &  2.21    \\
 $(4|44)$ & 2.72    & 3.78   & 6.45  & 5.37  &  1.08    & 1.60 & 2.55  & 2.36 & 4.76    & 7.02 & 11.03  &  7.82    \\
 $(1|55)$ & 1.19    & 1.71   & 3.29  & 2.80  &  0.25    & 0.42 & 0.58  & 0.79 & 0.91    & 1.50 &  2.08  &  2.24    \\
 $(5|55)$ & 2.41    & 3.85   & 6.78  & 6.50  &  1.03    & 1.69 & 2.74  & 2.54 & 4.57    & 7.39 & 12.17  &  9.07    \\
 $(1|66)$ & 0.72    & 1.58   & 2.75  & 2.81  &  0.16    & 0.35 & 0.47  & 0.63 & 0.55    & 1.19 &  1.66  &  1.95    \\
 $(6|66)$ & 2.86    & 4.37   & 6.76  & 8.24  &  1.18    & 1.85 & 2.79  & 3.08 & 5.21    & 8.14 & 12.08  & 11.28   \\
\hline\hline
\end{tabular}
}

  \caption{Speedup of the MD implementation of Coulomb 3-center 2-particle integrals in \code{LibintX} vs the reference OSHGPA implementation in \code{Libint}.$^a$}
  \label{table:md3}
  $^a$ SIMD instruction sets: I1 $\equiv$ scalar x86-64, I4 $\equiv$ AVX2, I8 $\equiv$ AVX512, I2 $\equiv$ NEON.
\end{table}

\subsection{4-center 2-particle integrals}

\Cref{table:md4} reports the performance of our MD implementation for 4-center 2-particle integrals.
Scalar implementation is never slower than the reference OSHGP code, despite the lower operation count of the latter, even in contracted cases due to the partial early contraction optimization.

With increasing angular momentum the relative SIMD performance decreases, due to BLAS GEMM (internally vectorized) accounting for a larger fraction of time.  Since non-SIMD bra transformation (unlike SIMD) can use BLAS GEMM, in cases where bra transform dominates, eg uncontracted $(55|xx)$ and $(66|xx)$ type integrals, non-SIMD code performs comparable or better than SIMD counterpart.  However, with increasing ket contraction this advantage fades as bra cost decreases in comparison.

The unexpectedly poor NEON performance in certain cases is due to GEMM not being used.
Apple Accelerate performs poorly for small kernels, such as the $|dd)$ ket, likely due to suboptimal problem size and the lack of the JIT API comparable to the Intel MKL.

Explicit vectorization does increase the memory required for the intermediates, especially the 2-index Hermite integrals.  For example, the $(hh|hh)$ integrals need $286^2=81796$ $[{\bf \tilde{p}}|{\bf \tilde{q}}]$ integrals per vector lane or over 2 MB per AVX2 core, which exceeds the 1MB L3 cache of a typical x86-64 core. This is especially noticable when multiple threads compete for memory bandwidth; this is illustrated in \cref{table:md4-6core} where the SIMD $(hh|hh)$ kernel is found to be nearly 2 times slower than the scalar counterpart.  Such cases are not typical; e.g., this example assumes that each core executes the same high-$L$ kernel and are unlikely to affect the overall real-world performance; however, if needed these issues are straightforward to resolve.

Performance for the 4-center integrals was also assessed against the established \code{Simint} library of Pritchard and Chow\cite{VRG:pritchard:2016:JCC} that emits SIMD instructions by vectorizing over the primitives in the vertical recurrence part of the OSHGP algorithm as well as by batching over multiple shellsets. For all integral classes and contraction degrees the present MD implementation outperforms \code{Simint} by a healthy margin. This occurs even for high contraction depth where the advantage of the OSHGP algorithm over the MD algorithm should be most pronounced, and the abundant data parallelism over primitives can be exploited most efficiently by \code{Simint}.



\begin{table}[ht!]
 \scalebox{1.0}{
\begin{tabular}{l|llll|llll|llll}
 \hline\hline
 & \multicolumn{4}{c|}{$\{K_\text{bra},K_\text{ket}\} = \{1, 1\}$}  & \multicolumn{4}{c|}{$\{K_\text{bra},K_\text{ket}\} = \{1, 10\}$}     &  \multicolumn{4}{c}{$\{K_\text{bra},K_\text{ket}\} = \{5, 10\}$} \\
 \hline
 $(l_\mathbf{a}l_\mathbf{b}|l_\mathbf{c}l_\mathbf{d})$     & I1 & I4   & I8 & I2  & I1 & I4   & I8 & I2 & I1 & I4  & I8 & I2    \\
 \hline
 $(00|00)$ & 4.57    & 7.36  & 8.87   & 5.24  & 4.23    & 5.42 & 5.84   & 3.97 & 3.92    & 5.37 & 5.55            & 3.53 \\
 $(10|00)$ & 10.69   & 26.02 & 26.14  & 11.76 & 5.48    & 9.48 & 10.08  & 5.38 & 4.27    & 7.26 & 7.77            & 4.54 \\
 $(11|00)$ & 6.93    & 16.44 & 17.63  & 6.82  & 3.71    & 7.83 & 9.17   & 4.30 & 2.81    & 5.76 & 7.19            & 3.26 \\
 $(11|11)$ & 2.31    & 6.07  & 5.76   & 2.64  & 1.30    & 4.58 & 5.23   & 1.90 & 1.03    & 3.52 & 4.61            & 1.55 \\
 $(22|00)$ & 3.58    & 5.98  & 6.49   & 3.19  & 2.45    & 4.88 & 4.88   & 2.45 & 1.82    & 3.63 & 3.86            & 1.88 \\
 $(22|11)$ & 2.25    & 4.20  & 6.26   & 2.73  & 1.52    & 4.71 & 7.19   & 3.05 & 1.22    & 3.79 & 5.90            & 2.48 \\
 $(22|22)$ & 3.00    & 3.81  & 6.05   & 1.53  & 2.44    & 3.09 & 5.00   & 1.26 & 2.09    & 2.66 & 4.35            & 1.09 \\
 $(33|00)$ & 1.99    & 2.93  & 3.53   & 1.48  & 1.68    & 3.29 & 4.14   & 1.70 & 1.31    & 2.50 & 3.20            & 1.31 \\
 $(33|11)$ & 2.19    & 3.10  & 4.04   & 2.09  & 1.59    & 4.19 & 5.89   & 2.89 & 1.24    & 3.27 & 4.84            & 2.29 \\
 $(33|33)$ & 3.65    & 4.26  & 5.78   & 3.16  & 2.32    & 2.69 & 3.67   & 5.47 & 1.93    & 2.27 & 2.99            & 5.00 \\
 $(44|00)$ & 2.19    & 2.71  & 3.02   & 1.99  & 2.43    & 4.75 & 5.51   & 3.33 & 2.02    & 3.72 & 4.43            & 2.77 \\
 $(44|11)$ & 2.03    & 2.40  & 2.78   & 1.88  & 1.54    & 3.97 & 4.99   & 2.97 & 1.20    & 3.03 & 3.90            & 2.34 \\
 $(44|44)$ & 3.51    & 4.33  & 7.12   & 4.54  & 1.85    & 2.68 & 3.87   & 4.20 & 1.53    & 2.21 & 3.13            & 3.65 \\
 $(55|00)$ & 1.93    & 1.58  & 1.76   & 1.38  & 2.47    & 3.53 & 4.10   & 2.81 & 1.97    & 2.77 & 3.03            & 2.32 \\
 $(55|11)$ & 1.96    & 1.40  & 1.72   & 1.32  & 2.80    & 2.82 & 3.27   & 1.78 & 2.10    & 2.09 & 2.29            & 1.31 \\
 $(55|55)$ & 5.49    & 5.40  & 8.84   & 5.79  & 2.15    & 2.58 & 4.01   & 4.55 & 1.59    & 1.92 & 2.97            & 3.68 \\
 $(66|00)$ & 2.06    & 1.91  & 1.89   & 1.39  & 2.43    & 3.92 & 4.13   & 2.77 & 1.88    & 2.94 & 2.97            & 2.14 \\
 $(66|11)$ & 1.70    & 1.48  & 2.02   & 1.19  & 1.72    & 2.33 & 3.14   & 1.29 & 1.24    & 1.67 & 2.01            & 0.87 \\
 $(66|66)$ & 14.37   & 13.04 & 20.88  & 9.18  & 3.36    & 3.80 & 6.14   & 5.79 & 1.80    & 1.96 & 3.43            & 3.77 \\
\hline\hline
\end{tabular}
}

  \caption{Speedup of the MD implementation of Coulomb 4-center 2-particle integrals in \code{LibintX} vs the reference OSHGP implementation in \code{Libint}.$^a$}
  \label{table:md4}
  $^a$ SIMD instruction sets: I1 $\equiv$ scalar x86-64, I4 $\equiv$ AVX2, I8 $\equiv$ AVX512, I2 $\equiv$ NEON.
\end{table}

\begin{table}[ht!]
 \begin{tabular}{c|ccc}
\hline\hline
$(l_\mathbf{a}l_\mathbf{b}|l_\mathbf{c}l_\mathbf{d})$ & \ $\{K_\text{bra},K_\text{ket}\} = \{1, 1\}$  & $\{K_\text{bra},K_\text{ket}\} = \{1, 10\}$     &  $\{K_\text{bra},K_\text{ket}\} = \{5, 10\}$ \\
 \hline
 $(00|00)$ & 0.82 & 0.96 & 0.93 \\
 $(11|11)$ & 0.62 & 0.36 & 0.34 \\
 $(22|22)$ & 0.87 & 0.90 & 0.85 \\
 $(33|33)$ & 1.02 & 1.00 & 0.93 \\
 $(44|44)$ & 1.12 & 0.86 & 0.72 \\
 $(55|55)$ & 1.84 & 1.39 & 1.27 \\
 $(66|66)$ & 1.36 & 1.29 & 1.20 \\
\hline\hline
\end{tabular}

 \caption{Slowdown of SIMD (I4) MD kernels for 4-center 2-particle integrals relative to the scalar (I1) counterparts while executing on all 6 cores of the Intel i7-9750H CPU. These examples illustrate the worst-case performance degradation due to the SIMD-driven memory pressure increase.}
 \label{table:md4-6core}
\end{table}

\begin{table}[ht!]
 \begin{tabular}{c|ccc}
\hline\hline $(l_\mathbf{a}l_\mathbf{b}|l_\mathbf{c}l_\mathbf{d})$         & \ $\{K_\text{bra},K_\text{ket}\} = \{1, 1\}$  & $\{K_\text{bra},K_\text{ket}\} = \{1, 10\}$     &  $\{K_\text{bra},K_\text{ket}\} = \{5, 10\}$      \\
\hline
 $(00|00)$ &   1.78  &   1.71   &   1.52         \\
 $(10|00)$ &   1.92  &   2.20   &   1.86         \\
 $(11|00)$ &   1.64  &   2.41   &   2.01         \\
 $(11|11)$ &   1.24  &   2.02   &   1.96         \\
 $(22|00)$ &   1.15  &   1.83   &   1.77         \\
 $(22|11)$ &   1.22  &   1.99   &   1.86         \\
 $(22|22)$ &   1.95  &   2.27   &   2.07         \\
 $(33|00)$ &   1.75  &   1.96   &   1.52         \\
 $(33|11)$ &   2.02  &   3.71   &   3.19         \\
 $(33|33)$ &   3.33  &   2.72   &   2.50         \\
 $(44|00)$ &   1.60  &   2.18   &   1.65         \\
 $(44|11)$ &   2.22  &   5.02   &   4.43         \\
 $(44|44)$ &   3.10  &   2.22   &   1.90         \\
 $(55|00)$ &   4.67  &   4.44   &   1.91         \\
 $(55|11)$ &   2.32  &   4.52   &   3.31         \\
 $(55|55)$ &   4.96  &   2.30   &   1.65         \\
\hline\hline
\end{tabular}

 \caption{Speedup of the MD implementation of Coulomb 4-center 2-particle integrals in \code{LibintX} vs the SIMD-optimized OSHGP implementation in \code{Simint} (Ref. \citenum{VRG:pritchard:2016:JCC}).$^a$}
 $^a$ Performance was measured on a single AVX2 core of the Intel i7-9750H CPU.
\end{table}

\section{Summary}
\label{sec:summary}

We reported an implementation of the McMurchie-Davidson evaluation scheme for 1- and 2-particle Gaussian AO integrals designed for efficient execution on modern CPUs with SIMD instruction sets. Like in our recent MD implementation for the GPUs\cite{VRG:asadchev:2023:JPCA,VRG:asadchev:2024:JCP} data parallelism is achieved by computing batches of shellsets of integrals at a time, rather than the one shellset at a time approach popular in the popular CPU-only integral engines like our own \code{Libint} engine.

By optimizing for the floating point instruction throughput rather than minimizing the number of operations, this approach achieves up to 50\% of the theoretical hardware peak FP64 performance for many common SIMD-equipped platforms (AVX2, AVX512, NEON), which translates to speedups of up to $30$ over the state-of-the-art one-shellset-at-a-time implementation of Obara-Saika-type schemes in \code{Libint} for a variety of primitive and contracted integrals.  In this context the relative GPU speedups are much more modest, e.g., 28, 46, and 83 rather than 592, 223, and 1172 for uncontracted $(11|00)$, $(22|22)$, $(66|66)$, respectively, when comparing V100 numbers from our prior work and i7-9750h numbers here.

As with our previous work, we rely on the standard C++ programming language --- such as the {\tt std::simd} standard library feature to be included in the 2026 ISO C++ standard --- without any explicit code generation to keep the code base small and portable. The implementation is part of the open source {\tt LibintX} library freely available at \url{https://github.com/ValeevGroup/libintx}. We invite interested parties to collaborate on the development of fundamental open-source high-performance quantum chemistry building blocks.

\begin{suppinfo}

Raw benchmark performance data, platform information.

\end{suppinfo}

\begin{acknowledgement}
This research was supported by the US Department of Energy, Office of Science, via award DE-SC0022327. The authors acknowledge Advanced Research Computing at Virginia Tech (https://arc.vt.edu/) for providing computational resources and technical support that have contributed to the results reported in this paper.
\end{acknowledgement}

\begin{appendices}


\end{appendices}

\bibliography{vrgrefs}

\providecommand{\latin}[1]{#1}
\makeatletter
\providecommand{\doi}
  {\begingroup\let\do\@makeother\dospecials
  \catcode`\{=1 \catcode`\}=2 \doi@aux}
\providecommand{\doi@aux}[1]{\endgroup\texttt{#1}}
\makeatother
\providecommand*\mcitethebibliography{\thebibliography}
\csname @ifundefined\endcsname{endmcitethebibliography}  {\let\endmcitethebibliography\endthebibliography}{}
\begin{mcitethebibliography}{26}
\providecommand*\natexlab[1]{#1}
\providecommand*\mciteSetBstSublistMode[1]{}
\providecommand*\mciteSetBstMaxWidthForm[2]{}
\providecommand*\mciteBstWouldAddEndPuncttrue
  {\def\EndOfBibitem{\unskip.}}
\providecommand*\mciteBstWouldAddEndPunctfalse
  {\let\EndOfBibitem\relax}
\providecommand*\mciteSetBstMidEndSepPunct[3]{}
\providecommand*\mciteSetBstSublistLabelBeginEnd[3]{}
\providecommand*\EndOfBibitem{}
\mciteSetBstSublistMode{f}
\mciteSetBstMaxWidthForm{subitem}{(\alph{mcitesubitemcount})}
\mciteSetBstSublistLabelBeginEnd
  {\mcitemaxwidthsubitemform\space}
  {\relax}
  {\relax}

\bibitem[Asadchev and Valeev(2023)Asadchev, and Valeev]{VRG:asadchev:2023:JPCA}
Asadchev,~A.; Valeev,~E.~F. High-{{Performance Evaluation}} of {{High Angular Momentum}} 4-{{Center Gaussian Integrals}} on {{Modern Accelerated Processors}}. \emph{J. Phys. Chem. A} \textbf{2023}, \emph{127}, 10889--10895\relax
\mciteBstWouldAddEndPuncttrue
\mciteSetBstMidEndSepPunct{\mcitedefaultmidpunct}
{\mcitedefaultendpunct}{\mcitedefaultseppunct}\relax
\EndOfBibitem
\bibitem[Asadchev and Valeev(2023)Asadchev, and Valeev]{VRG:asadchev:2023:JCTC}
Asadchev,~A.; Valeev,~E.~F. Memory-{{Efficient Recursive Evaluation}} of 3-{{Center Gaussian Integrals}}. \emph{J. Chem. Theory Comput.} \textbf{2023}, \emph{19}, 1698--1710\relax
\mciteBstWouldAddEndPuncttrue
\mciteSetBstMidEndSepPunct{\mcitedefaultmidpunct}
{\mcitedefaultendpunct}{\mcitedefaultseppunct}\relax
\EndOfBibitem
\bibitem[Asadchev and Valeev(2024)Asadchev, and Valeev]{VRG:asadchev:2024:JCP}
Asadchev,~A.; Valeev,~E.~F. 3-Center and 4-Center 2-Particle {{Gaussian AO}} Integrals on Modern Accelerated Processors. \emph{J. Chem. Phys.} \textbf{2024}, \emph{160}, 244109\relax
\mciteBstWouldAddEndPuncttrue
\mciteSetBstMidEndSepPunct{\mcitedefaultmidpunct}
{\mcitedefaultendpunct}{\mcitedefaultseppunct}\relax
\EndOfBibitem
\bibitem[Yokogawa \latin{et~al.}(2024)Yokogawa, Ito, Tsuji, Fujii, Suzuki, Nakano, and Kasagi]{VRG:yokogawa:2024:2TISCNWC}
Yokogawa,~N.; Ito,~Y.; Tsuji,~S.; Fujii,~H.; Suzuki,~K.; Nakano,~K.; Kasagi,~A. Parallel {{GPU Computation}} of {{Nuclear Attraction Integrals}} in {{Quantum Chemistry}}. 2024 {{Twelfth Int}}. {{Symp}}. {{Comput}}. {{Netw}}. {{Workshop CANDARW}}. Naha, Japan, 2024; pp 163--169\relax
\mciteBstWouldAddEndPuncttrue
\mciteSetBstMidEndSepPunct{\mcitedefaultmidpunct}
{\mcitedefaultendpunct}{\mcitedefaultseppunct}\relax
\EndOfBibitem
\bibitem[Boys(1950)]{VRG:boys:1950:PRSMPES}
Boys,~S.~F. Electronic Wave Functions. {{I}}. {{A}} General Method of Calculation for the Stationary States of Any Molecular System. \emph{Proc. R. Soc. Math. Phys. Eng. Sci.} \textbf{1950}, \emph{200}, 542--554\relax
\mciteBstWouldAddEndPuncttrue
\mciteSetBstMidEndSepPunct{\mcitedefaultmidpunct}
{\mcitedefaultendpunct}{\mcitedefaultseppunct}\relax
\EndOfBibitem
\bibitem[Sun(2015)]{VRG:sun:2015:JCC}
Sun,~Q. Libcint: {{An}} Efficient General Integral Library for {{Gaussian}} Basis Functions. \emph{J. Comput. Chem.} \textbf{2015}, \emph{36}, 1664--1671\relax
\mciteBstWouldAddEndPuncttrue
\mciteSetBstMidEndSepPunct{\mcitedefaultmidpunct}
{\mcitedefaultendpunct}{\mcitedefaultseppunct}\relax
\EndOfBibitem
\bibitem[Sun(2024)]{VRG:sun:2024:JCP}
Sun,~Q. The Updates in {{Libcint}} 6: {{More}} Integrals, {{API}} Refinements, and {{SIMD}} Optimization Techniques. \emph{J. Chem. Phys.} \textbf{2024}, \emph{160}, 174116\relax
\mciteBstWouldAddEndPuncttrue
\mciteSetBstMidEndSepPunct{\mcitedefaultmidpunct}
{\mcitedefaultendpunct}{\mcitedefaultseppunct}\relax
\EndOfBibitem
\bibitem[Obara and Saika(1986)Obara, and Saika]{VRG:obara:1986:JCP}
Obara,~S.; Saika,~A. Efficient Recursive Computation of Molecular Integrals over {{Cartesian Gaussian}} Functions. \emph{J. Chem. Phys.} \textbf{1986}, \emph{84}, 3963--3974\relax
\mciteBstWouldAddEndPuncttrue
\mciteSetBstMidEndSepPunct{\mcitedefaultmidpunct}
{\mcitedefaultendpunct}{\mcitedefaultseppunct}\relax
\EndOfBibitem
\bibitem[{Head-Gordon} and Pople(1988){Head-Gordon}, and Pople]{VRG:head-gordon:1988:JCP}
{Head-Gordon},~M.; Pople,~J.~A. A Method for Two-electron {{Gaussian}} Integral and Integral Derivative Evaluation Using Recurrence Relations. \emph{J. Chem. Phys.} \textbf{1988}, \emph{89}, 5777--5786\relax
\mciteBstWouldAddEndPuncttrue
\mciteSetBstMidEndSepPunct{\mcitedefaultmidpunct}
{\mcitedefaultendpunct}{\mcitedefaultseppunct}\relax
\EndOfBibitem
\bibitem[Pritchard and Chow(2016)Pritchard, and Chow]{VRG:pritchard:2016:JCC}
Pritchard,~B.~P.; Chow,~E. Horizontal Vectorization of Electron Repulsion Integrals. \emph{J. Comput. Chem.} \textbf{2016}, \emph{37}, 2537--2546\relax
\mciteBstWouldAddEndPuncttrue
\mciteSetBstMidEndSepPunct{\mcitedefaultmidpunct}
{\mcitedefaultendpunct}{\mcitedefaultseppunct}\relax
\EndOfBibitem
\bibitem[Weigend and Ahlrichs(2005)Weigend, and Ahlrichs]{VRG:weigend:2005:PCCP}
Weigend,~F.; Ahlrichs,~R. Balanced Basis Sets of Split Valence, Triple Zeta Valence and Quadruple Zeta Valence Quality for {{H}} to {{Rn}}: {{Design}} and Assessment of Accuracy. \emph{Phys. Chem. Chem. Phys.} \textbf{2005}, \emph{7}, 3297\relax
\mciteBstWouldAddEndPuncttrue
\mciteSetBstMidEndSepPunct{\mcitedefaultmidpunct}
{\mcitedefaultendpunct}{\mcitedefaultseppunct}\relax
\EndOfBibitem
\bibitem[McMurchie and Davidson(1978)McMurchie, and Davidson]{VRG:mcmurchie:1978:JCP}
McMurchie,~L.~E.; Davidson,~E.~R. One- and Two-Electron Integrals over Cartesian Gaussian Functions. \emph{J. Comp. Phys.} \textbf{1978}, \emph{26}, 218--231\relax
\mciteBstWouldAddEndPuncttrue
\mciteSetBstMidEndSepPunct{\mcitedefaultmidpunct}
{\mcitedefaultendpunct}{\mcitedefaultseppunct}\relax
\EndOfBibitem
\bibitem[Whitten(1973)]{VRG:whitten:1973:JCP}
Whitten,~J.~L. Coulombic Potential Energy Integrals and Approximations. \emph{J. Chem. Phys.} \textbf{1973}, \emph{58}, 4496--4501\relax
\mciteBstWouldAddEndPuncttrue
\mciteSetBstMidEndSepPunct{\mcitedefaultmidpunct}
{\mcitedefaultendpunct}{\mcitedefaultseppunct}\relax
\EndOfBibitem
\bibitem[Baerends \latin{et~al.}(1973)Baerends, Ellis, and Ros]{VRG:baerends:1973:CP}
Baerends,~E.~J.; Ellis,~D.~E.; Ros,~P. Self-Consistent Molecular {{Hartree}}---{{Fock}}---{{Slater}} Calculations i. {{The}} Computational Procedure. \emph{Chem. Phys.} \textbf{1973}, \emph{2}, 41--51\relax
\mciteBstWouldAddEndPuncttrue
\mciteSetBstMidEndSepPunct{\mcitedefaultmidpunct}
{\mcitedefaultendpunct}{\mcitedefaultseppunct}\relax
\EndOfBibitem
\bibitem[Dunlap(2000)]{VRG:dunlap:2000:PCCP}
Dunlap,~B.~I. Robust and Variational Fitting. \emph{Phys. Chem. Chem. Phys.} \textbf{2000}, \emph{2}, 2113--2116\relax
\mciteBstWouldAddEndPuncttrue
\mciteSetBstMidEndSepPunct{\mcitedefaultmidpunct}
{\mcitedefaultendpunct}{\mcitedefaultseppunct}\relax
\EndOfBibitem
\bibitem[Hohenstein \latin{et~al.}(2012)Hohenstein, Parrish, Sherrill, and Mart{\'i}nez]{VRG:hohenstein:2012:JCPa}
Hohenstein,~E.~G.; Parrish,~R.~M.; Sherrill,~C.~D.; Mart{\'i}nez,~T.~J. Communication: {{Tensor}} Hypercontraction. {{III}}. {{Least-squares}} Tensor Hypercontraction for the Determination of Correlated Wavefunctions. \emph{J. Chem. Phys.} \textbf{2012}, \emph{137}, 221101\relax
\mciteBstWouldAddEndPuncttrue
\mciteSetBstMidEndSepPunct{\mcitedefaultmidpunct}
{\mcitedefaultendpunct}{\mcitedefaultseppunct}\relax
\EndOfBibitem
\bibitem[Pierce \latin{et~al.}(2021)Pierce, Rishi, and Valeev]{VRG:pierce:2021:JCTC}
Pierce,~K.; Rishi,~V.; Valeev,~E.~F. Robust {{Approximation}} of {{Tensor Networks}}: {{Application}} to {{Grid-Free Tensor Factorization}} of the {{Coulomb Interaction}}. \emph{J. Chem. Theory Comput.} \textbf{2021}, \emph{17}, 2217--2230\relax
\mciteBstWouldAddEndPuncttrue
\mciteSetBstMidEndSepPunct{\mcitedefaultmidpunct}
{\mcitedefaultendpunct}{\mcitedefaultseppunct}\relax
\EndOfBibitem
\bibitem[Lehtola \latin{et~al.}(2020)Lehtola, Visscher, and Engel]{VRG:lehtola:2020:JCP}
Lehtola,~S.; Visscher,~L.; Engel,~E. Efficient Implementation of the Superposition of Atomic Potentials Initial Guess for Electronic Structure Calculations in {{Gaussian}} Basis Sets. \emph{J. Chem. Phys.} \textbf{2020}, \emph{152}, 144105\relax
\mciteBstWouldAddEndPuncttrue
\mciteSetBstMidEndSepPunct{\mcitedefaultmidpunct}
{\mcitedefaultendpunct}{\mcitedefaultseppunct}\relax
\EndOfBibitem
\bibitem[Schlegel and Frisch(1995)Schlegel, and Frisch]{VRG:schlegel:1995:IJQC}
Schlegel,~H.~B.; Frisch,~M.~J. Transformation between {{Cartesian}} and Pure Spherical Harmonic {{Gaussians}}. \emph{Int. J. Quantum Chem.} \textbf{1995}, \emph{54}, 83--87\relax
\mciteBstWouldAddEndPuncttrue
\mciteSetBstMidEndSepPunct{\mcitedefaultmidpunct}
{\mcitedefaultendpunct}{\mcitedefaultseppunct}\relax
\EndOfBibitem
\bibitem[Helgaker \latin{et~al.}(2000)Helgaker, J{\o}rgensen, and Olsen]{VRG:helgaker:2000:}
Helgaker,~T.; J{\o}rgensen,~P.; Olsen,~J. \emph{Molecular Electronic-Structure Theory}, 1st ed.; Helgaker/{{Molecular}} Electronic-Structure Theory; John Wiley \& Sons, Ltd: Chichester, UK, 2000\relax
\mciteBstWouldAddEndPuncttrue
\mciteSetBstMidEndSepPunct{\mcitedefaultmidpunct}
{\mcitedefaultendpunct}{\mcitedefaultseppunct}\relax
\EndOfBibitem
\bibitem[Ahlrichs(2006)]{VRG:ahlrichs:2006:PCCP}
Ahlrichs,~R. A Simple Algebraic Derivation of the {{Obara-Saika}} Scheme for General Two-Electron Interaction Potentials. \emph{Phys. Chem. Chem. Phys.} \textbf{2006}, \emph{8}, 3072\relax
\mciteBstWouldAddEndPuncttrue
\mciteSetBstMidEndSepPunct{\mcitedefaultmidpunct}
{\mcitedefaultendpunct}{\mcitedefaultseppunct}\relax
\EndOfBibitem
\bibitem[Neese(2022)]{VRG:neese:2022:JCC}
Neese,~F. The {{SHARK}} Integral Generation and Digestion System. \emph{J Comput Chem} \textbf{2022}, \emph{44}, 381--396\relax
\mciteBstWouldAddEndPuncttrue
\mciteSetBstMidEndSepPunct{\mcitedefaultmidpunct}
{\mcitedefaultendpunct}{\mcitedefaultseppunct}\relax
\EndOfBibitem
\bibitem[Ahlrichs(2004)]{VRG:ahlrichs:2004:PCCP}
Ahlrichs,~R. Efficient Evaluation of Three-Center Two-Electron Integrals over {{Gaussian}} Functions. \emph{Phys. Chem. Chem. Phys.} \textbf{2004}, \emph{6}, 5119\relax
\mciteBstWouldAddEndPuncttrue
\mciteSetBstMidEndSepPunct{\mcitedefaultmidpunct}
{\mcitedefaultendpunct}{\mcitedefaultseppunct}\relax
\EndOfBibitem
\bibitem[Peels and Knizia(2020)Peels, and Knizia]{VRG:peels:2020:JCTC}
Peels,~M.; Knizia,~G. Fast Evaluation of Two-Center Integrals over Gaussian Charge Distributions and Gaussian Orbitals with General Interaction Kernels. \emph{J. Chem. Theory Comput.} \textbf{2020}, \emph{16}, 2570--2583\relax
\mciteBstWouldAddEndPuncttrue
\mciteSetBstMidEndSepPunct{\mcitedefaultmidpunct}
{\mcitedefaultendpunct}{\mcitedefaultseppunct}\relax
\EndOfBibitem
\bibitem[{Williams-Young} \latin{et~al.}(2023){Williams-Young}, Asadchev, Popovici, Clark, Waldrop, Windus, Valeev, and De~Jong]{VRG:williams-young:2023:JCP}
{Williams-Young},~D.~B.; Asadchev,~A.; Popovici,~D.~T.; Clark,~D.; Waldrop,~J.; Windus,~T.~L.; Valeev,~E.~F.; De~Jong,~W.~A. Distributed Memory, {{GPU}} Accelerated {{Fock}} Construction for Hybrid, {{Gaussian}} Basis Density Functional Theory. \emph{J. Chem. Phys.} \textbf{2023}, \emph{158}, 234104\relax
\mciteBstWouldAddEndPuncttrue
\mciteSetBstMidEndSepPunct{\mcitedefaultmidpunct}
{\mcitedefaultendpunct}{\mcitedefaultseppunct}\relax
\EndOfBibitem
\end{mcitethebibliography}

\end{document}